\documentclass[12pt]{iopart}

\usepackage{amsfonts}
\usepackage{amssymb}
\usepackage{graphicx}
\usepackage{float}
\usepackage{color}
\usepackage{subfigure}
\newcommand{\Gstr}{G^\mathrm{ss}_\mathrm{str}(n,f)}
\newcommand{\Gform}{G_\mathrm{form}(n)}
\newcommand{\dxss}{\Delta x^\mathrm{ss}_l}
\newcommand{\dxsse}{\Delta x^\mathrm{ss}}

\begin{document}

\title{Unfolding kinetics of periodic DNA hairpins}

\author{Sandra Nostheide$^{\rm a}$, Viktor Holubec$^{\rm b}$, Petr
  Chvosta$^{\rm b}$ and Philipp Maass$^{\rm a}$}

\address{$^{\rm a}$ Fachbereich Physik, Universit\"at Osnabr\"uck,
  Barbarastra{\ss}e 7, 49076 Osnabr\"uck, Germany}

\address{$^{\rm b}$ Department of Macromolecular Physics, Faculty of
  Mathematics and Physics, Charles University, CZ-180~00~Praha, Czech
  Republic} 

\eads{\mailto{chvosta@kmf.troja.mff.cuni.cz},
  \mailto{philipp.maass@uni-osnabrueck.de}}

\date{\today}

\begin{abstract}
  DNA hairpin molecules with periodic base sequences can be expected
  to exhibit a regular coarse-grained free energy landscape (FEL) as
  function of the number of open base pairs and applied mechanical
  force. Using a commonly employed model, we first analyse for which
  types of sequences a particularly simple landscape structure is
  predicted, where forward and backward energy barriers between partly
  unfolded states are decreasing linearly with force. Stochastic
  unfolding trajectories for such molecules with simple FEL are
  subsequently generated by kinetic Monte Carlo simulations.
  Introducing probabilities that can be sampled from these
  trajectories, it is shown how the parameters characterising the FEL
  can be estimated. Already 300 trajectories, as typically generated
  in experiments, provide faithful results for the FEL parameters.
\end{abstract}


\section{Introduction}

Single-molecule experiments deliver previously unprecedented insights
into kinetics and thermodynamics of biophysical and biochemical
processes \cite{Ritort2006, Hormeno2006,Kumar2010}. One class of these
experiments comprises unfolding and folding of DNA (or other
biomolecules) under application of mechanical forces by optical or
magnetic tweezers, or an atomic force microscope \cite{Conroy2004,
  Liphardt2001, Collin2005, Greenleaf2005, Woodside2006a, Mossa2009,
  Manosas2009, Huguet2010, Engel2011, Danilowicz2003, Janshoff2000,
  Zhuang2003, Alessandrini2005}.  An important aim of these studies is
to obtain information on free energy differences (FEDs) between folded
and unfolded states, or, more generally, on the free energy landscape
(FEL) characterising both the levels of stable as well as metastable,
partly unfolded states and the energy barriers in between them.

Equilibrium methods are the easiest way to determine FEDs.  In these
the Boltzmann statistics is applied to the fraction of residence times
of the states at different values of the control parameters
\cite{Liphardt2001, Mossa2009, Manosas2009, Li2006, Alemany2012,
  Ribezzi2013}. However, for many molecules the barriers between
states turn out to be so large that an equilibrium is not achieved
within the available experimental time window \cite{Liphardt2001,
  Alemany2012, Cocco2003}.

An intriguing way to measure FEDs from out-of-equilibrium measurements
is made possible by utilising detailed and integral work fluctuation
theorems \cite{Esposito2010, Seifert2012}, as the Crooks fluctuation
theorem \cite{Crooks1999}, or its integral counterpart, the Jarzynski
equality \cite{Jarzynski1997}.  Analogous to the Boltzmann
distribution in equilibrium, these theorems hold true in an universal
way independent of microscopic details of the molecule and its
unfolding/folding dynamics.  Work in the fluctuation theorems refers
to the thermodynamic work under a certain driving protocol
$\lambda(t)$ of control variables during a time interval $[t_i,t_f]$,
as, for example, pulling a molecule with a force $f(t)$. Because of
the strong fluctuations in single molecule experiments, work values
$W$ become stochastic variables with a distribution $p(W)$.

The Crooks theorem states that work distributions $p(W)$ and $p_R(W)$
for a protocol $\lambda(t)$ and the associated reversed protocol
$\lambda_R(t)=\lambda(t_i+t_f-t)$, respectively, are related according
to $p(W)/p_R(-W)=\textrm{e}^{(W-\Delta F)/k_{\rm B}T}$
\cite{Crooks1999}, where $\Delta F=F_f-F_i$ is the difference in
(equilibrium) free energies $F_{i,f}$ of the molecule in the
macrostates specified by the control variables
$\lambda_{i,f}=\lambda(t_{i,f})$, and $k_{\rm B}T$ is the thermal
energy.  In cases where the Crooks theorem \cite{Crooks1999} can not
be used, e.~g., because the work for the reverse process can not be
obtained with sufficient statistics \cite{Engel2011, Hummer2001, 
  Harris2007}, the Jarzynski equality offers an alternative way to the
FED. This theorem states that $\langle \textrm{e}^{-W/k_{\rm
    B}T}\rangle= \int \textrm{d}W p(W) \textrm{e}^{-W/k_{\rm
    B}T}=\textrm{e}^{-\Delta F/k_{\rm B}T}$ \cite{Jarzynski1997}
and has been applied successfully in various cases
\cite{Danilowicz2003, Harris2007, Liphardt2002, Ritort2002}.  In
general, however, the application of the Jarzynski equality turns out
to be difficult, because $\langle \textrm{e}^{-W/k_{\rm B}T}\rangle$
is dominated by rare trajectories with work values $W<\Delta F$
\cite{Jarzynski1997, Ritort2002, Zuckerman2002, Gore2003}. A possible
solution is to extend measured histograms of (typical) work values to
the tail regime $W\ll\Delta F$ by fitting to theoretical
predictions. To this end, some generic behaviour in the tail regime
needs to be assumed and attempts have been made recently in this
direction \cite{Engel2009,Nickelsen2011,Palassini2011}.  Even under
some assumption of the functional form of the tail, the extraction of
reliable free energy estimates from the measured data requires care,
because one needs to correctly take into account the dependence of
biasing effects on the number of measurements (single unfolding
trajectories) \cite{Ritort2002, Zuckerman2002, Gore2003,
  Palassini2011, Wood1991, Fox2003, Shirts2003, Ritort2008}.

If the statistics is not sufficient to allow reasonable application of
the nonequilibrium fluctuation theorems, an alternative route for
determining the FED, and more generally, the FEL, is to analyse the
kinetics of unfolding and/or folding, and to compare it with
predictions from theories. This requires to introduce parameters that
are specific for the molecule under investigation. For example, it was
shown \cite{Mossa2009} that one can extract the FEL out of first
rupture force distributions. By considering a two-state DNA folder
these distributions for both the unfolding and folding process are
connected to the distributions of the survival probabilities in the
folded and unfolded state. In the case of constant loading/unloading
rates, a clever representation \cite{Mossa2009} of the survival
probabilities allows one to determine energetic barriers and the FED
between the two states.
 
It is clear that such methods heavily rely on a good theoretical
description of the kinetics of DNA unfolding and folding. These are
commonly based on coarse-grained approaches, where molecular states
are specified by the number $n$ of open base pairs (bps). Models have
been established in the past \cite{Woodside2006a, Cocco2003,
  Manosas2006, SantaLucia1998, Lubensky2002, Manosas2005, Huguet2009,
  Messieres2011} that faithfully describe the free energies of these
states and their variation with an externally applied force $f$ in
terms of a FEL $G(n,f)$. An important task is to optimise parameters
in such models by suitable measurements, and to determine their
variation under change of the environment, such as salt concentration,
pH value or temperature of the solvent.

So far analytical treatments concentrated on two-state folders. In
experiments often long sequences are considered which exhibit many
minima in the FEL associated with partly unfolded states. An advantage
of these sequences is that one unfolding/folding process involves many
transitions, each of them giving information on the FEL. On the other
hand, theoretical treatments of the unfolding and folding kinetics of
multistate folders are difficult because each transition between
successive minima is characterised by distinct barriers and energy
differences between the involved states. It is thus useful to consider
periodic DNA sequences where transitions between states are
characterised by the same energetic parameters.
In this work we will model the unfolding kinetics of such periodic DNA
sequences and investigate whether it is possible to obtain reliable
information on the FEL based on a small number of unfolding
trajectories. Using standard modelling, we first show that simple
regular FELs can be obtained for periodic sequences, which keep their
characteristics over a wide range of pulling forces. KMC simulations
are then applied to generate unfolding trajectories for these
sequences, as surrogate for experimental ones. By fitting survival and
persistence probabilities obtained from these trajectories to
analytical results, we demonstrate that already about 300 trajectories
can give good estimates of FEL parameters.

\section{Free energy landscapes (FELs)}
\label{sec:FEL_general} 

In a coarse-grained approach, unfolding of DNA (and other molecules)
is often described by thermally activated transitions between minima
of a FEL $G(n,f)$, which is expressed as a function of the number $n$
of sequentially opened bps and the applied stretching force $f$. It
can be decomposed as \cite{Mossa2009, Engel2011, Manosas2006}
\begin{equation} 
G(n,f)=\Gform+\Gstr -f \dxss\,,
\label{eq:g}
\end{equation}
where the force-independent ''formation'' part $\Gform$ quantifies the
free energy release upon breaking of $n$ bps, and the part $-f \dxsse$
describes the energetic preference of unfolded states for larger $f$;
$\dxsse$ is the length of the unfolded single-stranded (ss) part. The
force-dependent ''stretching'' part $\Gstr$ takes into account the
decrease of entropy of the ssDNA when increasing the end-to-end
distances of the unfolded parts \cite{Mossa2009}.  Models borrowed
from polymer physics are commonly used to specify $\Gstr$.  For the
most prominent examples, the freely jointed chain (FJC)
\cite{Smith1996} and the worm-like chain (WLC) model
\cite{Bustamante1994a}, we give details in \ref{sec:FEL_app}.

For $\Gform$ we use the nearest neighbour (NN) model \cite{Zimm1964,
  Tinoco1962}
\begin{equation}
\Gform = \sum_{\mu=n+1}^{n_{\rm tot}-1} \epsilon_{\mu,\mu+1} + 
G_{\rm loop}\left(1-\delta\left(n,n_{\rm tot}\right)\right)\,,
\label{eq:gform}
\end{equation}
where $\epsilon_{\mu,\mu+1}$ is the interaction energy between
adjacent bases (see \ref{sec:FEL_app} for an example).  This model has
been applied successfully in various experiments \cite{Woodside2006a,
  Mossa2009, Cocco2003, Manosas2006, Huguet2009, Messieres2011,
  Talukder2011,Gross2011}, and its parameters have been continuously
improved. We take the ones recently reported in \cite{Huguet2010} with
the parameter values listed in the table~\ref{tab:NNBP} in
\ref{sec:FEL_app}.

At zero force, the free energy $G(n,f=0)=\Gform$ of long unbranched
hairpin molecules without any additional structures such as, e.~g.,
interior loops, increases monotonically with $n$, with varying local
rise $\Delta
G_\mathrm{form}(n)=G_\mathrm{form}(n+1)-G_\mathrm{form}(n)$.
Considering unzipping experiments with a constant loading rate $r$,
corresponding to a force increasing linearly with time,
\begin{equation}
f(t) = f_0 + rt\,,
\label{eq:force}
\end{equation}
the FEL gets tilted and local minima (''states'') develop at values
of $n$, where $\Delta G_\mathrm{form}(n)$ changes significantly,
i.~e.\ where $\Delta G_\mathrm{form}(n-1)$ and $\Delta
G_\mathrm{form}(n)$ differ strongly. Once the tilting is large enough
that the level of these minima becomes comparable to the energy of the
completely folded state, thermally activated transitions from one
minimum to the next drive the molecule into the unfolded state.  In
general, each of these transitions is characterised by distinct
forward and backward barriers, which also evolve differently in time.
This implies that theoretical treatments become increasingly difficult
with larger number of states.

With the goal to make a theoretical treatment of the kinetics
tractable, we are looking for periodic sequences, where the forward
and backward barriers decrease linearly with force and where the
levels of the states decrease linearly with $f n_\alpha$.  Here
$n_\alpha$ denotes the particular $n$ values of the states $\alpha$.
To be specific, we write for the forward barriers $\Delta(f)$, the
backward barriers $\Delta'(f)$, and the state free energies
$G_{\alpha}(f) \equiv G_{n_{\alpha}}(f)$: \numparts
\begin{eqnarray}
\label{eq:FEL_features}
\label{eq:FEL_features_a}
\Delta(f) & = \Delta_0 - \Delta_1 f\,, \\
\label{eq:FEL_features_b}
\Delta'(f) & = \Delta_0' - \Delta_1' f\,, \\
\label{eq:FEL_features_c}
G_{\alpha}(f) & = \alpha (g_0 - g_1 f )\,.
\end{eqnarray}
\endnumparts
Because $G_{\alpha+1}(f) - G_\alpha(f) = \Delta(f) - \Delta' (f)$, it
holds $g_0 = \Delta_0 - \Delta_0 '$ and $g_1 = \Delta_1 - \Delta_1 '$,
that means the landscape is specified by four independent parameters.
It is sufficient that
equations~(\ref{eq:FEL_features_a})-(\ref{eq:FEL_features_c}) are
obeyed in the range where the energy levels start to become comparable
to the one of the folded state, that means when for the first time
transitions can take place.  Whether, on the basis of
equation~(\ref{eq:g}), periodic DNA sequences satisfying
equations~(\ref{eq:FEL_features_a})-(\ref{eq:FEL_features_c}) can be
found will be discussed in the next section.

\section{Landscapes of periodic DNA sequences}
\label{sec:FEL}

We first analyse unbranched DNA hairpin molecules that have periodic
bp sequences without interior loops.  For these molecules,
equations~(\ref{eq:FEL_features_a})-(\ref{eq:FEL_features_c}) can be
fulfilled over a sufficiently wide force range if the periodicity
length $L$ of the units is small enough.  It turns out that the minima
are not very pronounced and the barriers are just a few multiples of
the thermal energy $k_\mathrm{B}T$.  Accordingly, FEDs of these
sequences should be measurable also from equilibrium occupancy
statistics \cite{Liphardt2001, Mossa2009, Manosas2009, Li2006,
  Alemany2012, Ribezzi2013}.  When considering interior loops attached
to the basic units, the requirements imposed by
equations~(\ref{eq:FEL_features_a})-(\ref{eq:FEL_features_c}) can be
well fulfilled with typically high barriers between the minima.
Additional parameters specifying the change of the FEL by the interior
loops \cite{SantaLucia1998, Zuker2003} then enter the description. We
notice that they are less accurate than those defining the bp
interactions \cite{Huguet2010} (see the table~\ref{tab:NNBP} in
app.~\ref{sec:FEL_app}).

\subsection{Hairpin molecules without interior loops}
\label{sec:mol_noloop}

\begin{figure}[b]
\begin{center} 
  \includegraphics[scale=0.47]{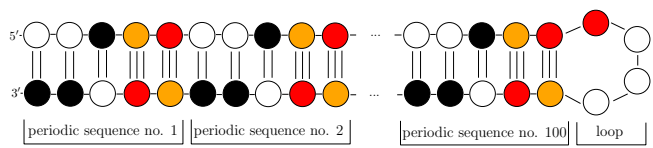}
\end{center}
\caption{Sketch of a periodic, random DNA hairpin structure with a
  periodicity length of $L=5$~bps. The bases are coloured as follows:
  adenine (white), cytosine (orange), guanine (red) and thymine
  (black).}
\label{fig:periodic_structure}
\end{figure}

A double-stranded (ds) DNA with periodicity length $L=5$~bps and an
arbitrarily chosen sequence in the period is sketched in
figure~\ref{fig:periodic_structure}.
Such hairpin molecules can, in principle, be branched or form other
structural elements.  To exclude such features we rely on folding
predictions of the program Mfold \cite{SantaLucia1998, Zuker2003}.
FELs for unbranched hairpin molecules, calculated according to
equations~(\ref{eq:g}) and (\ref{eq:gform}), in general do not satisfy
the requirements in
equations~(\ref{eq:FEL_features_a})-(\ref{eq:FEL_features_c}).
Starting from the folded state, minima and barriers first form
regularly in the landscape when $f$ is increased up to a certain
value.  However, by further increasing $f$, in between them new minima
and barriers can form. This means that the overall structure of the
regular landscape is not stable, but one regular type goes over to
another one.  The ''new barrier effect'' leads to force intervals
where equations~(\ref{eq:FEL_features_a})-(\ref{eq:FEL_features_c})
are satisfied, but jumps of $\Delta(f)$, $\Delta'(f)$ and
$G_\alpha(f)$ appear when going from one interval to the next one.
Another possibility that the regular structure gets modified is that
the positions of minima and saddle points are shifting.  This
''shifting effect'' leads to force intervals where
equations~(\ref{eq:FEL_features_a})-(\ref{eq:FEL_features_c}) are
satisfied, but jumps in the slopes $\Delta_1$, $\Delta_1'$ and $g_1$
appear when going from one interval to the next.  For long periodicity
lengths, the new barrier effect is difficult to avoid, even for
sequences that are specifically optimised in order to circumvent the
problem.  On the other hand, for sufficiently short periodicity
lengths ($L \lesssim 7$), the new barrier effect is usually not
occurring.  The shifting effect, however, affects the regular
behaviour also for small periodicity lengths. To avoid it, sequences
have to be optimised.

\begin{figure}[tb]
\subfigure
{\label{fig:rank11_FEL}
\includegraphics[scale=0.32,angle=-90]{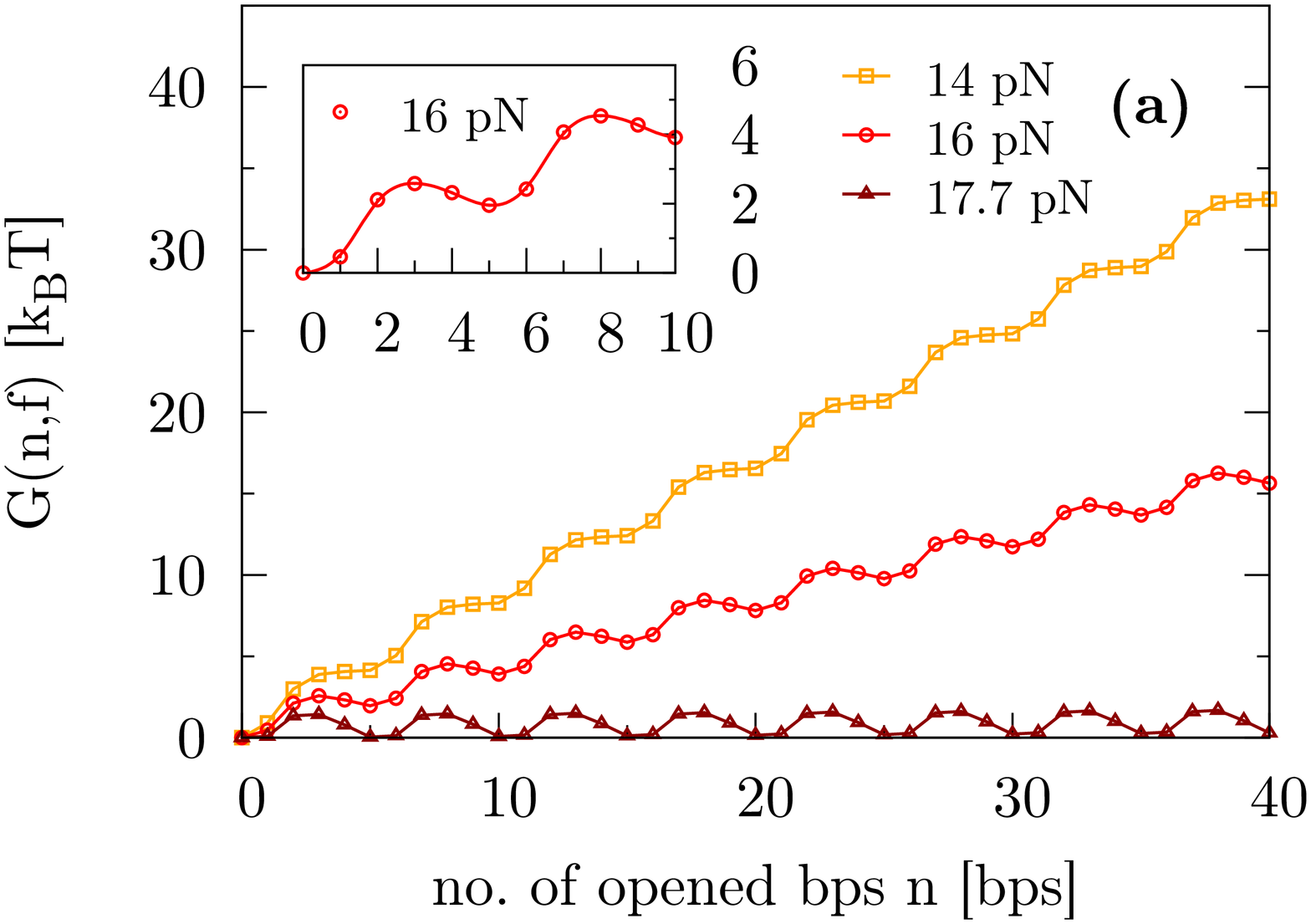}}
\subfigure
{\label{fig:rank11_FELparameters}
\includegraphics[scale=0.32,angle=-90]{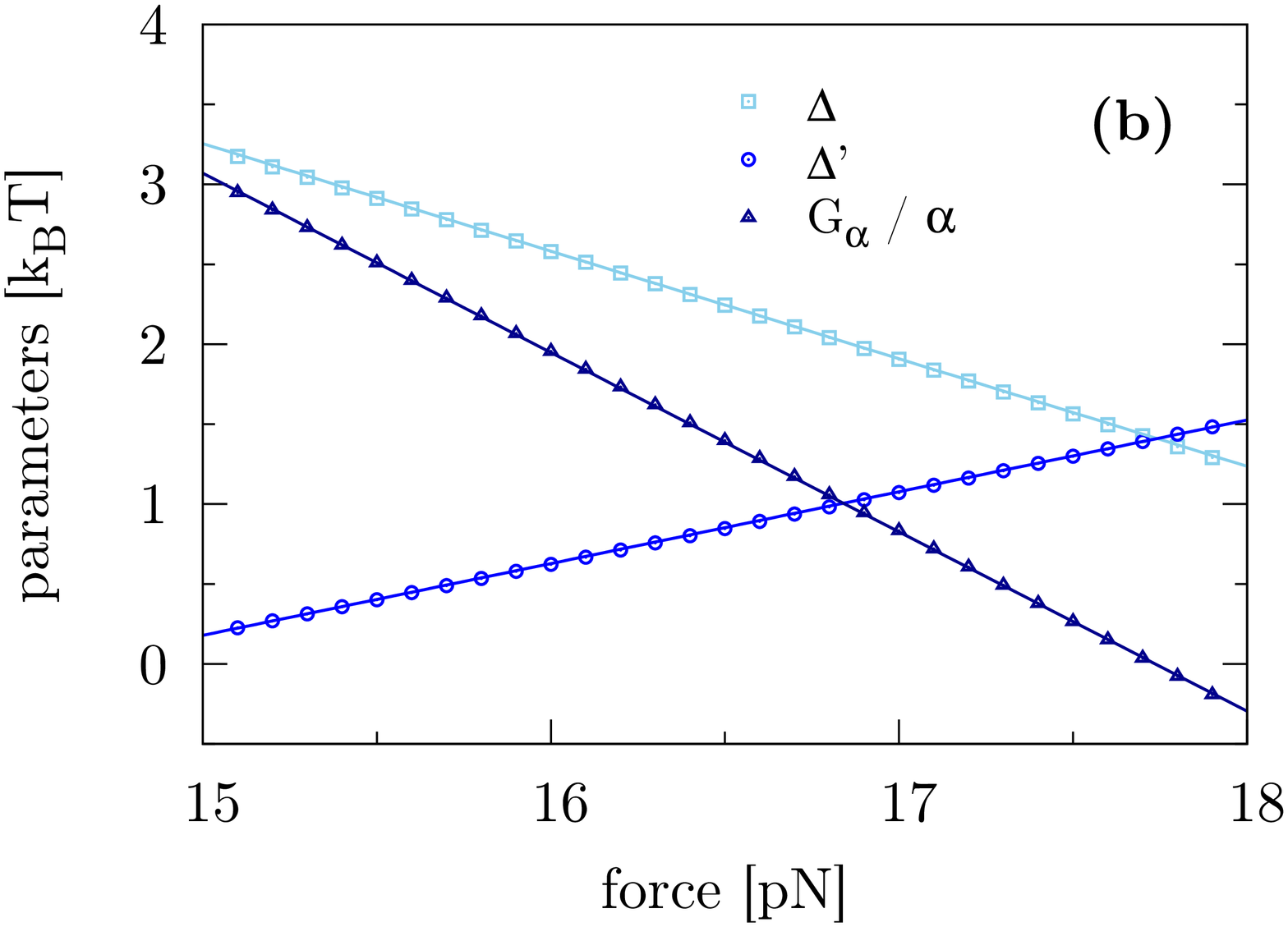}}
\caption{(a) FEL for the periodically continued sequence~I for three
  magnitudes of force. The sequence has TGCAA as basic unit and no
  interior loops. The inset shows a zoom of the first two barriers for
  $16$~pN. (b) Forward barriers $\Delta$, backward barriers $\Delta'$,
  and (order-normalised) state energies $G_\alpha/\alpha$ as a
  function of force for the sequence I. The regression lines show the
  linear behaviour according to
  equations~(\ref{eq:FEL_features_a})-(\ref{eq:FEL_features_c}).}
\label{fig:rank11}
\end{figure}

An optimisation among all $4^5$ sequences for $L=5$~bps with respect
to equations~(\ref{eq:FEL_features_a})-(\ref{eq:FEL_features_c}) gives
the smallest regression coefficients for the sequence consisting of
TGCAA [see figure~\ref{fig:rank11_FELparameters}], which we refer to
as sequence I.  The intercept and slope parameters of the linear
functions for this sequence I are given in the
table~\ref{tab:parameters}.  The corresponding FEL is depicted in
figure~\ref{fig:rank11_FEL} for three force values $f=14$~pN (minima
have not yet formed), $f=16$~pN (minima have formed, levels increase
with $n$), and $f=17.7$~pN (equal level of the minima).  The inset
shows two barriers zoomed out from the FEL for $f=16$~pN.  Because
these barriers are merely $2.8\,k_\mathrm{B}T$ high, they can be
easily surmounted at room temperature and equilibrium methods should
be applicable.

\begin{table}[b]
  \caption{Energetic parameters in 
    equations~(\ref{eq:FEL_features_a})-(\ref{eq:FEL_features_c}) for
    the three sequences I-III. (I:
    Optimised sequence with TGCAA as basic unit, without interior
    loops; II: Basic units with 10 A bases paired with T, separated by
    interior loops; III: Basic units with mixed types of bases of 
    $16$~bps length, separated by interior loops.}
\begin{indented}
\lineup
\item[]
\begin{tabular}{c c c c c}
\br
 parameter & unit & sequence I & sequence II & sequence III \\ \mr
 $\Delta_0$  & $k_\mathrm{B}T$ & \m 13.35 & \m 26.61 & \m 41.53 \\
 $\Delta_1$  & $k_\mathrm{B}T \over \mathrm{pN}$ & \m \0 0.67 & \0 \m 1.78 & \m \0 1.97 \\
 $\Delta_0'$ & $k_\mathrm{B}T$ & \0 $-6.55$ & \0 $-0.13$ & $-20.91$ \\ 
 $\Delta_1'$ & $k_\mathrm{B}T \over \mathrm{pN}$ & \0 $-0.45$   & \0 $-0.79$ & \0 $ -2.27$ \\ 
 $g_0$       & $k_\mathrm{B}T$ & \m 19.90 & \m 26.74 & \m 62.45 \\ 
 $g_1$       & $k_\mathrm{B}T \over \mathrm{pN}$ & \0 \m 1.12 & \0 \m 2.57 & \0 \m 4.23\\
\br
\end{tabular}
\label{tab:parameters}
\end{indented}
\end{table}

\subsection{Hairpin molecules with interior loops}
\label{sec:mol_loop}

\begin{figure}[b]
\subfigure
{\label{fig:L10_allA_FEL}
  \includegraphics[scale=0.32,angle=-90]{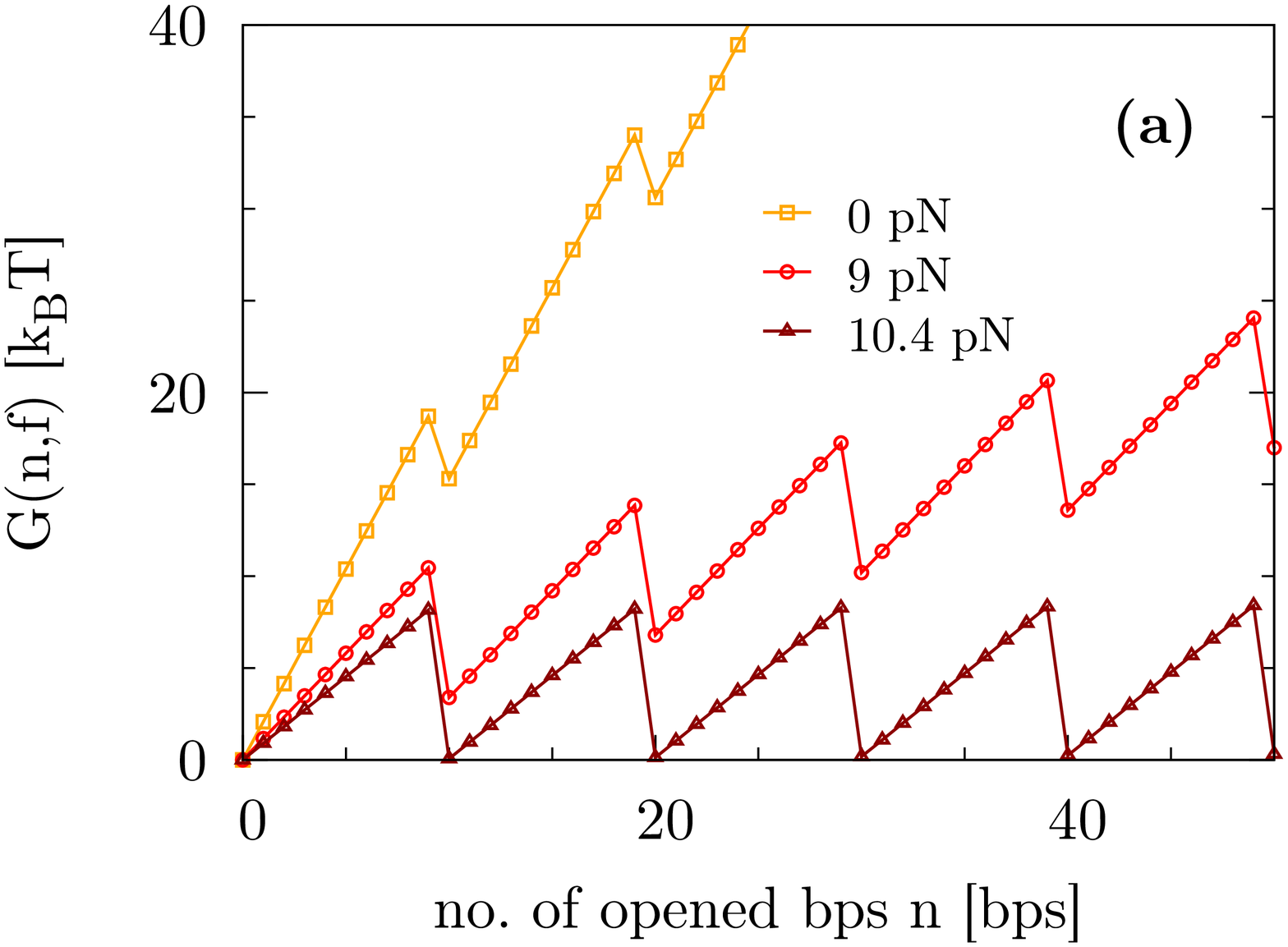}}
\subfigure 
{\label{fig:L10_allA_FELparameters}
  \includegraphics[scale=0.32,angle=-90]{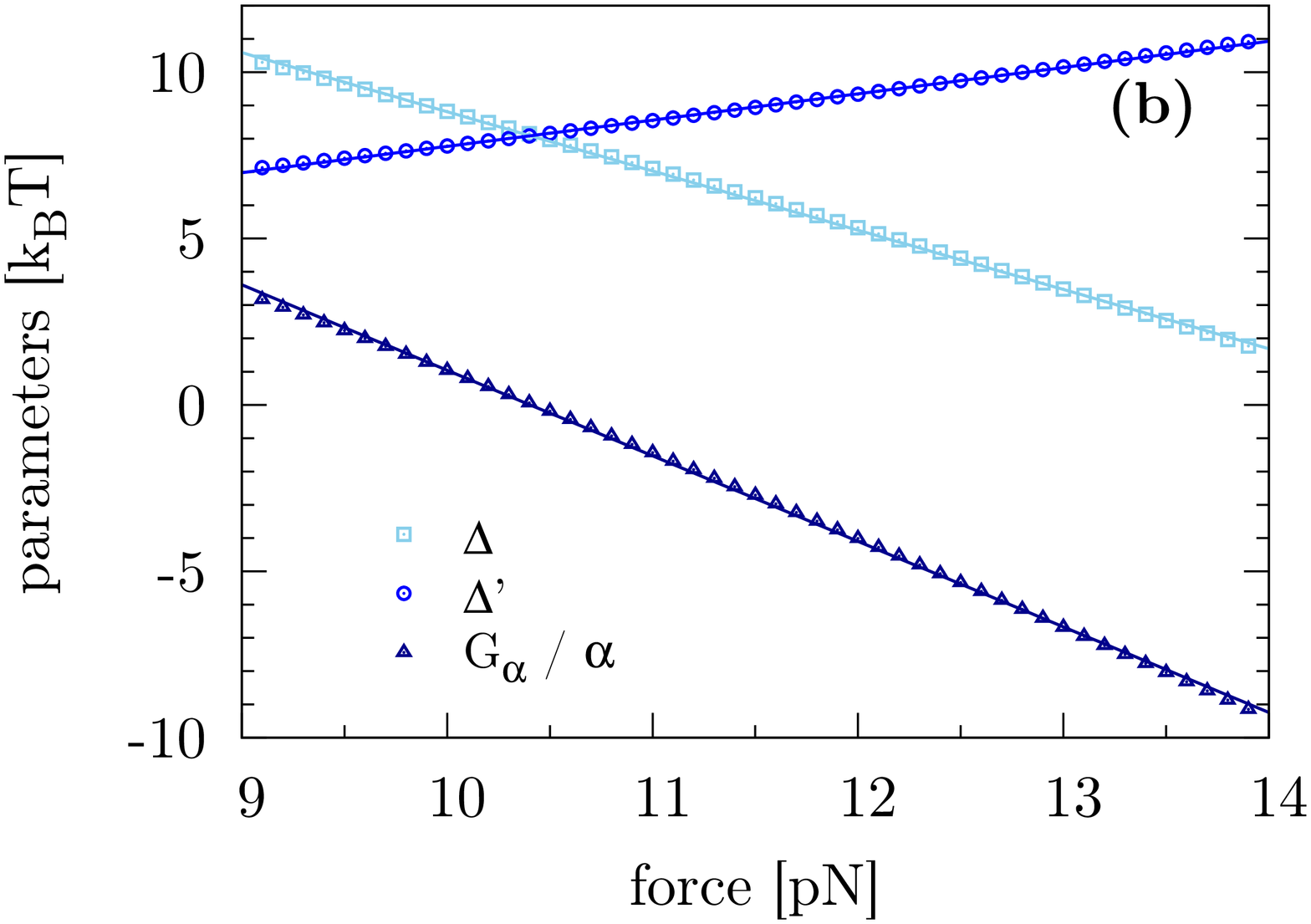}}
\caption{(a) FEL of the sequence II for three magnitudes of force.
  The sequence is composed of basic units each consisting 10 bases A
  and complementary bases T, and an interior loop, where TTT is
  attached to the A bases and CCC to the T bases.  (b) Forward
  barriers $\Delta$, backward barriers $\Delta'$, and
  (order-normalised) state energies $G_\alpha / \alpha$ as a function
  of force for the sequence II. The regression lines show the linear
  behaviour according to
  equations~(\ref{eq:FEL_features_a})-(\ref{eq:FEL_features_c}).}
\label{fig:L10_allA}
\end{figure}

For longer basic units,
equations~(\ref{eq:FEL_features_a})-(\ref{eq:FEL_features_c}) can be
satisfied by attaching interior loops to each unit.  In this case even
equal type of bases in the ds parts, e.~g. $\rm{AAA}\ldots$ paired
with $\rm{TTT}\ldots$, yield local minima in the FEL.  This is because
the ssDNA in the interior loops has more flexibility, which leads to
an entropy rise and accordingly a drop in $G(n,f)$ associated with
additional $G_{\rm loop}$ terms in equation~(\ref{eq:gform}).  Such
sequences with equal type of bases in the ds parts could be
particularly useful to determine improved $G_{\rm loop}$ values in
single-molecule experiments.

Let us consider a ds part consisting of bases A and complementary
bases T, and an interior loop with TTT attached to the A bases and CCC
attached to the T bases. Care was taken that Mfold
\cite{SantaLucia1998, Zuker2003} predicts a linear folding structure
for the molecule. The corresponding loop contribution to the free
energy $G_{\rm loop}$ is 3.24~kJ/mol. Barrier heights can be regulated
by varying the length of the ds part.  For example, for 10~bps in the
ds part, we find a barrier height of $8.1\,k_\mathrm{B}T$ at
$f=10.4$~pN, where all minima are at equal level.  For a sequence with
20 bps the corresponding force is $f=12.4$~pN and the barrier height
$9.7\,k_\mathrm{B}T$. We refer to the sequence with 10 bps as sequence
II. Its FEL is shown in figure~\ref{fig:L10_allA_FEL} for three force
values.  Figure \ref{fig:L10_allA_FELparameters} demonstrates that it
follows the linear behaviour according to
equations~(\ref{eq:FEL_features_a})-(\ref{eq:FEL_features_c}).  The
respective intercept and slope parameters of these functions are given
in the table~\ref{tab:parameters}.

\begin{figure}[t]
\begin{center}
  \includegraphics[scale=0.4,angle=-90]{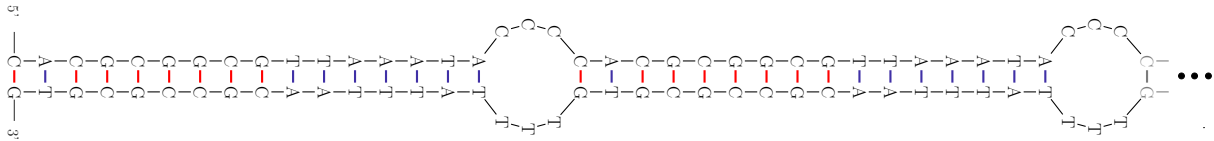}
\end{center}
\caption{DNA hairpin molecule with sequence III. One basic unit
  consists of a ds part with mixed types of bases, and a loop of ssDNA
  attached to it.}
\label{fig:structure_bestmola}
\end{figure}

\begin{figure}[t]
  \subfigure 
{\label{fig:branch_3xC3xT_FEL}
\includegraphics[scale=0.32,angle=-90]{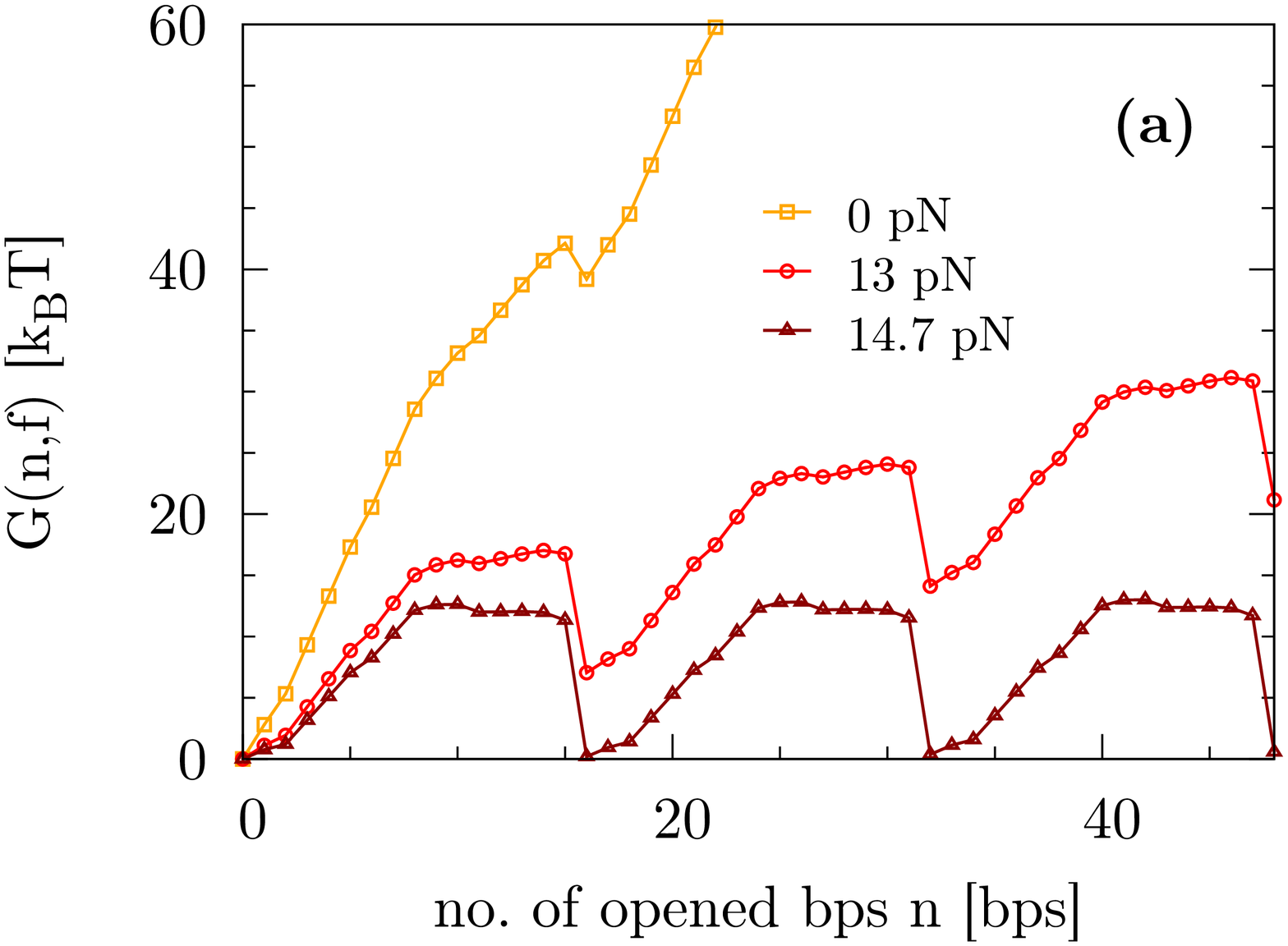}}
\subfigure
{\label{fig:branch_3xC3xT_FELparameters}
  \includegraphics[scale=0.32,angle=-90]{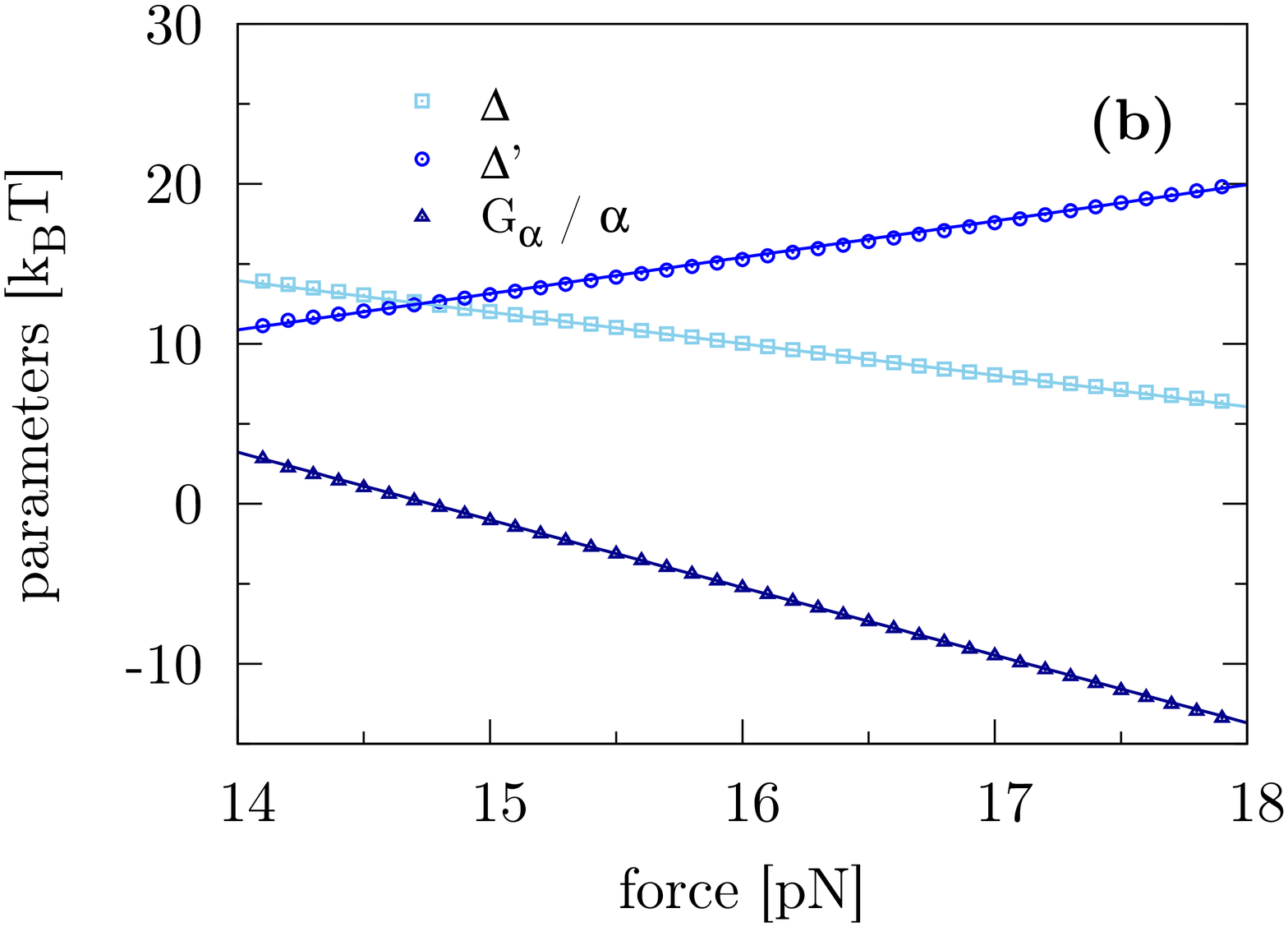}}
\caption{(a) FEL of the sequence III in
  Fig.~\ref{fig:structure_bestmola} for three force values. (b)
  Forward barriers $\Delta$, backward barriers $\Delta'$, and
  (order-normalised) state energies $G_\alpha/\alpha$ as a function of
  force for the sequence III. The regression lines show the linear
  behaviour according to
  equations~(\ref{eq:FEL_features_a})-(\ref{eq:FEL_features_c}).}
\label{fig:FEL_bestmol}
\end{figure}

To give an example for mixed types of bases in the ds part, we take
one of the branches of the triple-branch molecule investigated earlier
in \cite{Engel2011,Alemany2012} and attach to it the same loop as for
the molecule considered above with equal type of bases.  Only the
order is reversed, that means the CCC bases are located in the strand
closer to the 5' end. We refer to this sequence as sequence III.  The
respective $G_{\rm loop}$ value is $3.23$~kJ/mol. Figure
\ref{fig:structure_bestmola} depicts the molecule, and its FEL is
displayed in figure~\ref{fig:branch_3xC3xT_FEL} for three force
values. As shown in figure~\ref{fig:branch_3xC3xT_FELparameters}, it
follows equations~(\ref{eq:FEL_features_a})-(\ref{eq:FEL_features_c}).
The intercept and slope parameters of the linear functions are listed
in the table~\ref{tab:parameters}.  The barrier height is
$12.6\,k_\mathrm{B}T$ at $f=14.7$~pN, where all minima are at equal
level.

\section{Kinetics and Monte Carlo simulations}
\label{sec:kinetics}

According to standard transition rate theory \cite{Bell1978}, the rate
for surmounting the forward barrier is
\begin{eqnarray} 
\lambda(f) = \nu \;  \exp\left[- \beta \Delta(f)\right]\,,
\label{eq:gamma}
\end{eqnarray}
where $\nu$ is an attempt frequency and $\beta=1/k_\mathrm{B}T$. A
corresponding equation applies to the backward transition rate
$\lambda '(f)$ with the backward barrier $\Delta '(f)$ in the
exponential. Replacing $f$ by $t$ via equation~(\ref{eq:force}), these
rates become \numparts
\label{eq:transition_rates}
\begin{eqnarray}
\label{eq:forward_rate}
\lambda(t) & = \nu \; \exp[-\beta (\Delta_0-\Delta_1 f_0)] 
\; \exp(\beta \Delta_1 r t)\,,\\
\label{eq:backward_rate}
\lambda'(t) & =  \lambda(t) \; \exp[\beta (g_0-g_1 f_0)] \; 
\exp(- \beta g_1 r t).
\end{eqnarray}
\endnumparts
For a molecule with $m$ periodic units, the probabilities
$p_\alpha(t)$, $\alpha=0,\dots,m$, of being in state $\alpha$
evolve according to \numparts
\label{eq:pme}
\begin{eqnarray}
\label{eq:pme_first}
\frac{\textrm{d} p_0(t)}{\textrm{d}t} 
& = - \lambda(t) \; p_0(t) + \lambda '(t) \; p_1(t)\,,\\
\label{eq:pme_middle}
\frac{\textrm{d} p_\alpha(t)}{\textrm{d}t} 
& = - \lambda(t) \; \left[p_\alpha(t) - p_{\alpha-1}(t) \right] \\ \nonumber
& \phantom{{} = {}} {} + \lambda '(t) \; 
\left[p_{\alpha+1}(t) -p_\alpha(t) \right], \quad \alpha = 1, \ldots, m-1\,,\\
\label{eq:pme_last}
\frac{\textrm{d} p_m(t)}{\textrm{d}t} 
& = \lambda(t) \; p_{m-1}(t) - \lambda'(t) \; p_m(t)\,.
\end{eqnarray}
\endnumparts
While for a two-level system ($m=1$) an explicit solution is given in
\cite{Brey1991,Subrt_2007}, analytical treatments for larger $m$
become increasingly more difficult. Under neglect of backward
transitions, a solution for arbitrary $m$ is given in
\cite{Holubec2012}. Generally, the cumulative rate
\begin{equation}
\eqalign{\Lambda(t,t_0) & = 
  \int_{t_0}^t \textrm{d}t' \, \lambda (t') \nonumber \\
&{}=\frac{ \nu \; \exp[-\beta (\Delta_0-\Delta_1 f_0)]}{\beta \Delta_1 r }
\left[ \exp(\beta \Delta_1 r t) - \exp(\beta \Delta_1 r t_0) \right]}
\end{equation}
enters the solutions [as well as $\Lambda'(t,t_0)$]. With respect to
the goal to identify the parameters in
equations~(\ref{eq:FEL_features_a})-(\ref{eq:FEL_features_c}), it is
convenient to introduce the probability $\Psi_0(t)$ that, when
starting in the folded state ($\alpha=0$), no transition occurs until
time $t$. This is given by
\begin{eqnarray}
\Psi_0(t) = \exp[-\Lambda (t,0)]\,.
\label{eq:psi0}
\end{eqnarray} 
Because only the forward rate enters this expression, fitting to
experimental data should allow one to extract the two parameters
$\Delta_0$ and $\Delta_1$.  In order to have access to the two
remaining parameters $\Delta_0'$ and $\Delta_1'$ (or $g_0$ and $g_1$)
in equations~(\ref{eq:FEL_features_a})-(\ref{eq:FEL_features_c}), we
furthermore introduce the probability $\Psi_1(t)$ that, when starting
in the folded state ($\alpha=0$), at least one forward, but no
backward transition occurs until time $t$:
\begin{equation}
  \Psi_1(t) = \int_0^t 
  \textrm{d}t_1 \exp[-\Lambda' (t,t_1)] 
  \lambda(t_1) \exp[-\Lambda (t_1,0)]\,.
\label{eq:psi1}
\end{equation}
The term $\lambda(t_1) \exp[-\Lambda (t_1,0)]$ is the probability that
the first transition occurs at a time $t_1$, and the term
$\exp[-\Lambda' (t,t_1)]$ is the probability that thereafter no
backward transition takes place. For any loading rate, $\Psi_1(t)$
goes through a maximum and approaches a finite value in the long-time
limit. Since backward transitions become the less likely the larger
$r$, the maximum can become unnoticeable at high loading rates.

In order to test the procedure, we perform Monte Carlo simulations of
the stochastic process with the parameter values listed in the
table~\ref{tab:parameters}, which we will henceforth refer to as the
``true values''. In all cases we consider molecules with $m=4$
periodic units. The probabilities $\Psi_0(t)$ and $\Psi_1(t)$ are
sampled from a set of $N$ unfolding trajectories.  The simulations can
be performed in an efficient manner by using the First Reaction Time
Algorithm (FRTA) \cite{Einax2009,Holubec2011}. Starting from a
antecedent transition at time $t_0$, one generates a random attempt
time $t$ for a successive transition from a uniformly distributed
number $\eta$ in the unit interval by
\begin{eqnarray}
\Lambda(t,t_0) = -\ln \left(1-\eta\right)\,,
\label{eq:prob_ttime}
\end{eqnarray}
which yields 
\begin{eqnarray}
t = \frac{1}{\beta \Delta_1\eta} 
\ln \left( \exp \left(\beta \Delta_1 \eta t_0 \right) - 
\frac{\beta \Delta_1 \eta}{\nu \textrm{e}^{-\beta\left(\Delta_0- \Delta_1
      f_0 \right)}} 
\ln \left(1-\eta \right) \right)\,.
\end{eqnarray}
Analogously, an attempt time $t'$ for a backward transition is
generated from a further random number $\eta'$ (with $\Delta_0'$ and
$\Delta_1'$ replacing $\Delta_0$ and $\Delta_1$).  The transition
following the previous transition at time $t_0$ then takes place at
time $\min(t,t')$, and the corresponding forward or backward
transition is carried out.

\begin{figure}[t]
\subfigure
{\label{fig:typical_trajectory_r5}
\includegraphics[scale=0.32,angle=-90]{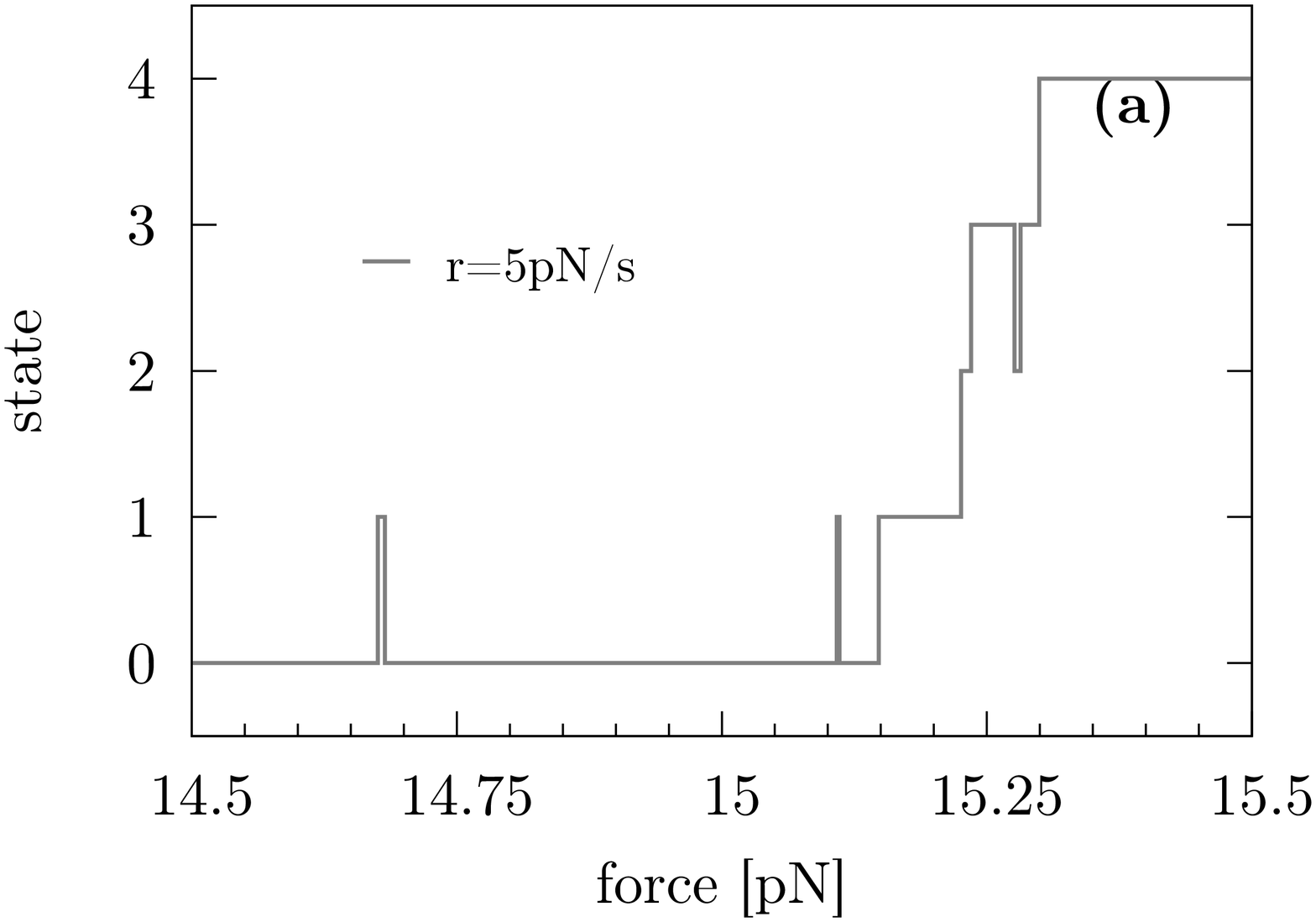}}
\subfigure
{\label{fig:typical_trajectory_r50}
\includegraphics[scale=0.32,angle=-90]{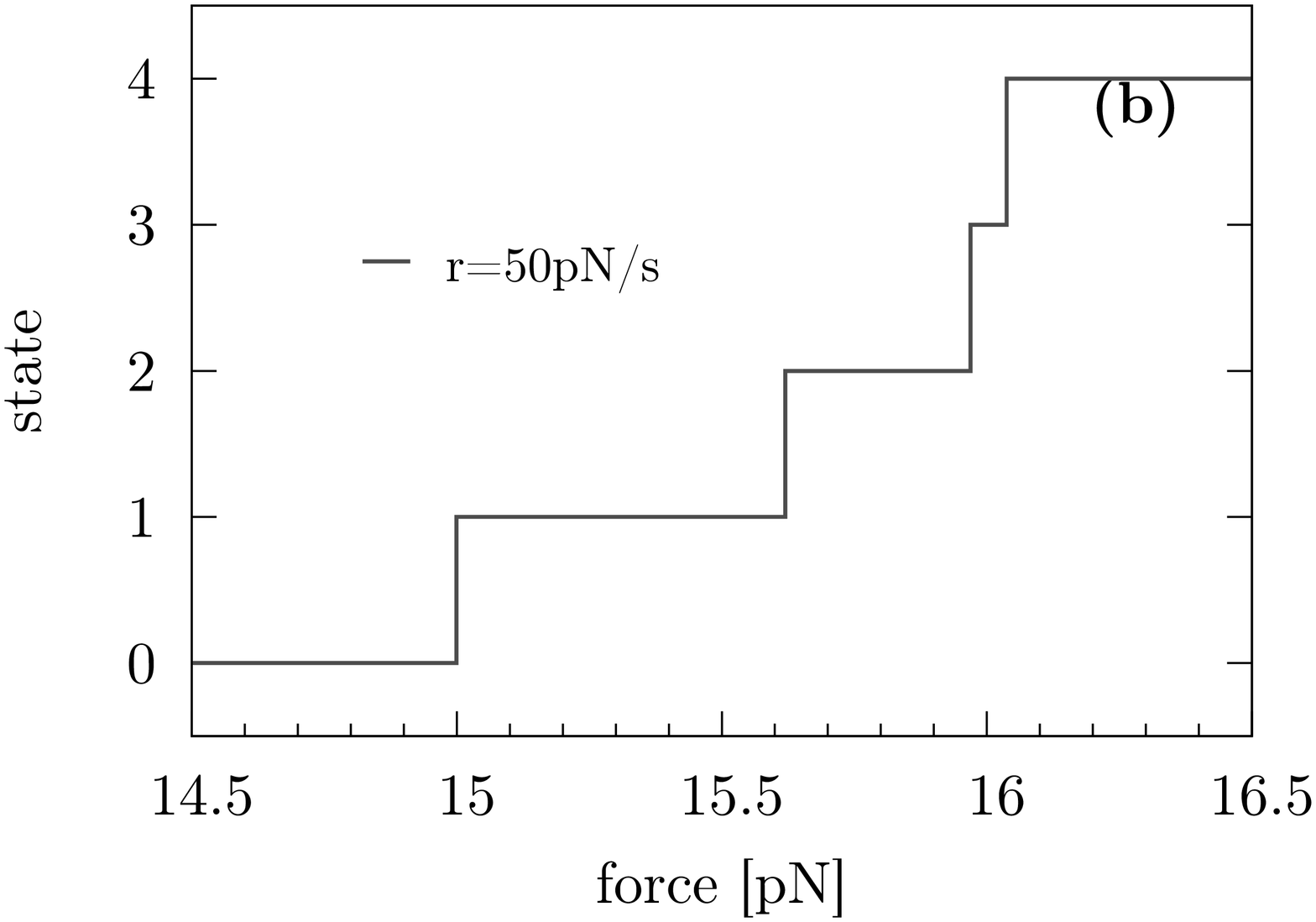}}
\caption{Molecular state as a function of the time-dependent force
  $f(t) = f_0 +rt$ for two different loading rates $r$ in the relevant
  time (force) regime of unfolding [for smaller or larger times, the
  molecule is in the folded ($\alpha=0$) or unfolded ($\alpha=m$)
  state, respectively].}
\label{fig:typical_trajectory}
\end{figure}

In the following we give a detailed presentation of the procedure
applied to the sequence III with mixed types of bases in the basic
units and interior loops. The other sequences, I and II, are treated
in the same way. Let us note that for the sequence I without interior
loops, a description of the unfolding kinetics based on a jump
dynamics is not really appropriate, because the barriers between
minima are too small (see the discussion in
section~\ref{sec:mol_noloop}).  Keeping this in mind, we nevertheless
include an analysis of this sequence for completeness.

Figure~\ref{fig:typical_trajectory} shows typical simulated
trajectories for (a) $r =5$~pN/s and (b) $r =50$~pN/s, plotted as a
function of the time-dependent force $f(t)=f_0+rt$.  As expected from
figure~\ref{fig:branch_3xC3xT_FEL}, the first transitions occur at
around $f=15$~pN, and then the molecule rapidly unfolds.  Moreover,
backward transitions become the less frequent the higher the loading
rate $r$ is.  In figure~\ref{fig:typical_trajectory_r5} three backward
transitions take place before reaching the unfolded state, while in
figure~\ref{fig:typical_trajectory_r50} no backward transition is
seen. If backward transitions are rare before complete unfolding,
information on the backward barriers is hardly contained in sampled
$\Psi_1(t)$. As a consequence, estimates for $\Delta_0'$ and
$\Delta_1'$ will become less accurate for larger $r$. A good signature
for sufficient statistics is that the long-time limit of $\Psi_1(t)$
is between zero and one (see also the discussion further down).

\begin{figure}[tb]
\subfigure
{\label{fig:psi0_curve}
\includegraphics[scale=0.31,angle=-90]{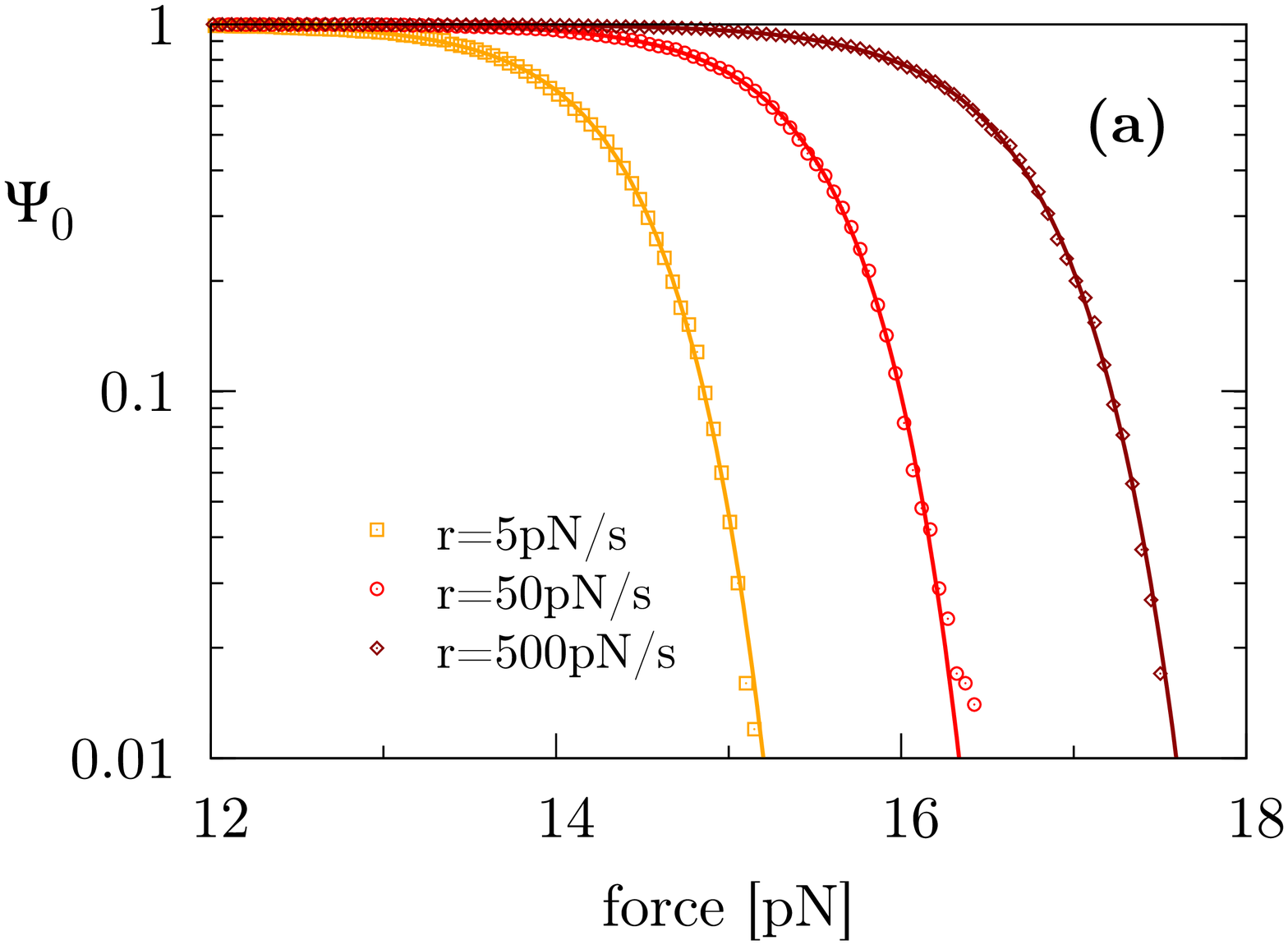}}
\subfigure
{\label{fig:psi1_curve}
\includegraphics[scale=0.31,angle=-90]{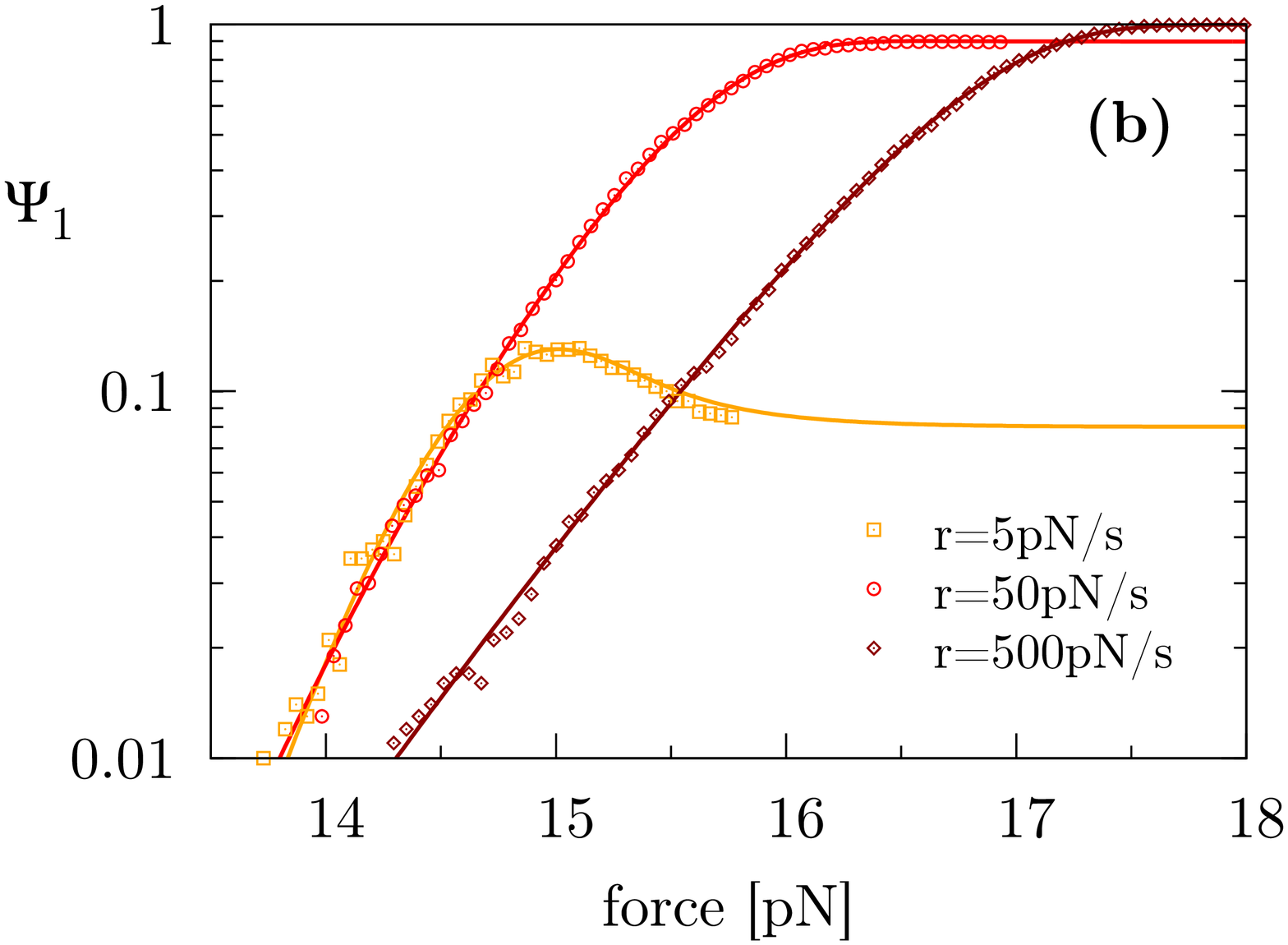}}
\caption{(a) $\Psi_0$ and (b) $\Psi_1$ as function of the
  time-dependent force $f=f(t)=f_0+rt$. Symbols refer to KMC
  simulations of $N=1000$ trajectories for three different loading
  rates. The solid lines are fits of equations~(\ref{eq:psi0}) and
  (\ref{eq:psi1}) to the data using the Levenberg-Marquardt
  algorithm.}
\label{fig:psi_curves}
\end{figure}

\begin{figure}[t]
\includegraphics[scale=0.63,angle=-90]{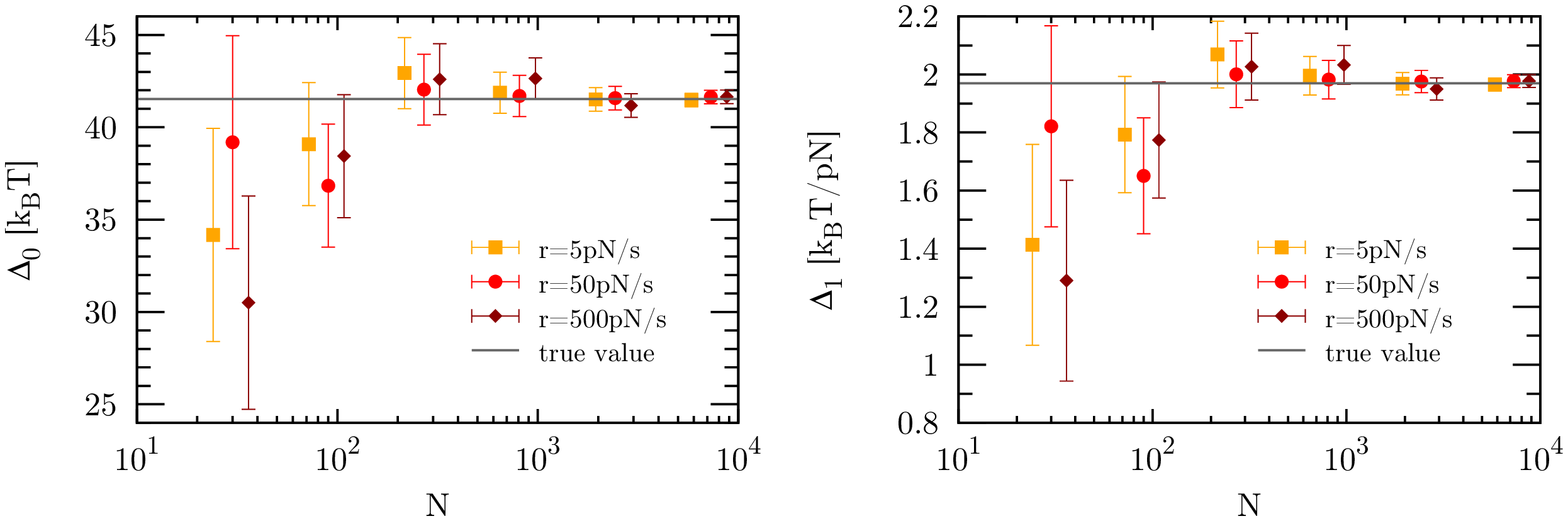}
\caption{Estimates of $\Delta_0$ (left panel) and $\Delta_1$ (right
  panel) based on the analysis of $N$ simulated unfolding trajectories
  for three different loading rates; $N$ values for $r=5$~pN/s and
  $r=500$~pN/s have been slightly shifted for better visibility.}
\label{fig:psi0_parameters}
\end{figure}

\begin{figure}[t]
\includegraphics[scale=0.63,angle=-90]{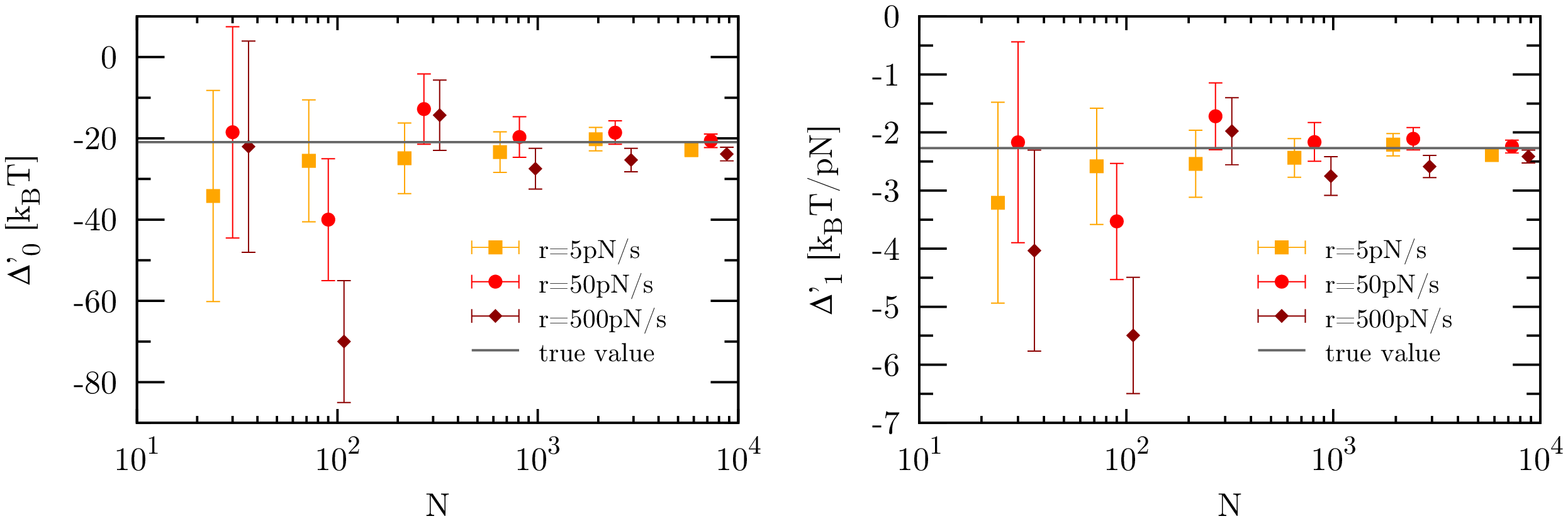}
\caption{Estimates of $\Delta_0'$ (left panel) and $\Delta_1'$ (right
  panel) based on the analysis of $N$ simulated unfolding trajectories
  for three different loading rates; $N$ values for $r=5$~pN/s and
  $r=500$~pN/s have been slightly shifted for better visibility.}
\label{fig:psi1_parameters}
\end{figure}

Figure~\ref{fig:psi0_curve} shows $\Psi_0$ sampled from $N=1000$
trajectories, for three different loading rates.  Nonlinear fitting of
equation~(\ref{eq:psi0}) to the data sets with the help of the
Levenberg-Marquardt algorithm \cite{nr} yields the solid
lines.\footnote{In the algorithm, in addition to the estimated
  $\Psi_j(t)$ values, $j=0,1$, their standard deviations $\sigma_j(t)$
  need to be given. These can be obtained in the simulations by
  repeating the procedure for estimating $\Psi_j(t)$ for a large
  number of sets of $N$ trajectories. They show a pronounced
  dependence on $t$, with a maximum in the ''unfolding regime'' (cf.\
  figure~\ref{fig:typical_trajectory}). Fortunately, taking
  time-independent $\sigma_j$ has a negligible effect on the best fit
  parameters $\Delta_j$. This means that constant $\sigma_j$ can be
  used in an experimental realisation, where only one set of $N$
  trajectories is available.}  The simulated data follow closely the
predicted behaviour and the fit parameters are shown in
figure~\ref{fig:psi0_parameters} as a function of the number of
generated trajectories.  To obtain 1000 trajectories in
single-molecule experiments is already an ambitious task, but not out
of reach.  Already for about 300 trajectories the fitted $\Delta_0$
and $\Delta_1$ are very close to the true values. In simulations,
larger numbers of trajectories can be generated. As is evident from
figure~\ref{fig:psi0_parameters}, the fit parameters converge to the
true values for $N\to\infty$.

With respect to applications, it is encouraging that already for 300
trajectories the fitted $\Delta_0$ and $\Delta_1$ deviate by a few
percent only, even when starting the fit at guessed input values which
have a deviation of 40 percent from the true parameters.  However, in
an experiment it would be difficult to estimate the standard
deviation. In figure~\ref{fig:psi0_parameters} the error bars were
obtained by repeating many times the analysis for a given number $N$
of trajectories.

As outlined above, we can now in a second step determine the backward
barriers $\Delta'_0$ and $\Delta'_1$ by analysing $\Psi_1(t)$, after
fixing the forward barriers to their fitted values.  Fits of
equation~(\ref{eq:psi1}) to the KMC data are shown in
figure~\ref{fig:psi1_curve}. The procedure is fully analogous to that
of determining $\Delta_0$ and $\Delta_1$, and
figure~\ref{fig:psi1_parameters} displays the corresponding results.
In fact, with respect to the quality of the fitting, we obtained
equivalent findings.  In particular, already 300 trajectories are
sufficient to yield parameter estimates that deviate by just a few
percent from the true values.

Analogous analyses have been performed for the sequences I and II.
The corresponding results, as obtained from 1000 trajectories, are
listed in the table~\ref{tab:parameter_results}, together with the
values from figures~\ref{fig:psi0_parameters} and
\ref{fig:psi1_parameters} for the sequence III. For the sequence I, it
was only possible to extract $\Delta_0$ and $\Delta_1$, because the
backward barriers are so small that backward transitions quickly occur
after the first forward transition. As a result, $\Psi_1(t)$ is nearly
zero for all $t$. The same holds true for the sequence II when using
the small unfolding rate $r=5$~pN/s. To overcome this problem, one
could generalise the treatment by introducing probabilities that at
least $k>1$ backward transitions occur up to time $t$. Note that
$1-\psi_1(t)$ is the probability that up to time $t$ at least one
backward transition takes place. The complexity of analytical
expressions for such probabilities, however, increases with $k$, which
makes a fitting to sampled data difficult in practise. As discussed in
section~\ref{sec:mol_noloop}, application of equilibrium methods would
be more suitable for sequences with small energetic barriers.  In all
cases, where a determination was possible, good estimates of the
energetic parameters in
equations~(\ref{eq:FEL_features_a})-(\ref{eq:FEL_features_c}) are
obtained. As explained above, the estimates become less accurate at
high $r$, where $\Psi_1(t)$ approaches values close to one at long
times. To get good estimates, one can tune the loading rate, such that
the long-time limit $\Psi_1(t)$ is about one half.

\begin{table}[t]
  \caption{Estimates of the energetic parameters in
    equations~(\ref{eq:FEL_features_a})-(\ref{eq:FEL_features_c}) 
    for the sequences I-III based on a sampling of $\Psi_0(t)$ and $\Psi_1(t)$ 
    from $N=1000$ unfolding trajectories. Results are given for three 
    different loading rates $r$. 
    An asterisk indicates that a
    determination of $\Delta_0'$ and $\Delta_1'$ from $\Psi_1(t)$ was not
    possible (see discussion in main text).}
\begin{indented}
\lineup
\item[]
\begin{tabular}{c c c c c c c c}
  \br
  \multicolumn{1}{l}{sequence} 
& $r$ & $\Delta_0$ & $\Delta_1$ & $\Delta_0'$ &  $\Delta_1'$ & $g_0$ & $g_1$ \\ 
\multicolumn{1}{r}{unit} & pN/s &  $k_\mathrm{B}T$ & $k_\mathrm{B}T \over 
\mathrm{pN}$ & $k_\mathrm{B}T$ & $k_\mathrm{B}T \over \mathrm{pN}$ 
&  $k_\mathrm{B}T$ & $k_\mathrm{B}T \over \mathrm{pN}$ \\ \mr
  & true & 13.35 & 0.67 & -6.55 & -0.45 & 19.90 & 1.12 \\
  & 5 & 13.34 & 0.65 & * & * & * & *\\
  \raisebox{1.5ex}[-1.5ex]{I} & 50 & 13.35 & 0.66 & * & * & * & *\\
  & 500 & 13.28 & 0.64 & * & * & * & *\\ \mr
  & true & 26.61 & 1.78 & -0.13 & -0.79 & 26.74 & 2.57 \\
  & 5 & 26.63 & 1.79 & * & * & * & * \\
\raisebox{1.5ex}[-1.5ex]{II} & 50 & 26.76 & 1.79 & -0.05 & -0.78 & 26.81 & 2.57 \\
 & 500 & 26.59 & 1.77 & 0.22 & -0.74 & 26.37 & 2.51 \\ \mr
 & true & 41.53 & 1.97 & -20.91 & -2.27 & 62.45 & 4.23 \\
 & 5 & 42.09 & 2.01 & -24.66 & -2.51 & 66.75 & 4.52\\
\raisebox{1.5ex}[-1.5ex]{III} & 50 & 42.41 & 2.03 & -17.43 & -2.04 & 59.84 & 4.07\\
 & 500 & 39.31 & 1.83 & - 15.96 & -1.97 & 55.27 & 3.80 \\ \br
\end{tabular}
\label{tab:parameter_results}
\end{indented}
\end{table}

\section{Summary and Conclusions}
\label{sec:conclusions}

Based on a commonly used model, we showed that the FEL for different
types of DNA hairpin molecules with periodic base sequences is
expected to change in a regular manner when applying an external
mechanical force $f$.  This regular change manifests itself in forward
and backward barriers between successive states that decrease linearly
with $f$ over a wide range of forces, covering regimes where molecules
completely unfold.  Using KMC simulations, stochastic unfolding
trajectories were generated as surrogate for experimental ones. With
the aim to identify the parameters of the FEL, two probabilities were
introduced: The probability $\Psi_0(t)$ that a molecule remains in the
folded state until time $t$, and the probability $\Psi_1(t)$ that
until time $t$, a molecule undergoes at least one forward but no
backward transition. These probabilities can be sampled from the
trajectories, and theoretical calculations give rather simple
expressions for them, allowing one to determine the FEL parameters by
a nonlinear fitting procedure. We demonstrated that already about 300
unfolding trajectories are sufficient to obtain good parameter
estimates. Best results are obtained if the loading rate is suitably
chosen, such that the long-time limit of $\Psi_1(t)$ is neither close
to zero nor close to one. For small barriers between states of the
order of a few $k_\mathrm{B}T$, such rates may become too high to be
realisable in experiment.

From a theoretical point of view, the possibility to generate simple
functional forms of FELs as function of the number $n$ of open base
pairs and applied force $f$ provides an interesting basis for future
studies, where exact results could be obtained for kinetic and
energetic properties. Analytical results for work distributions will
be of particular interest to gain a deeper understanding of tail
regimes.

With respect to applications, our findings provide a promising means
to obtain improved values of free energies of interior loops. By
attaching loops at regular spacings to homogeneous double strands,
periodic sequences can be synthesised and analysed. Taking several
periods, good counting statistics should be achieved.

\ack Support of this work by the Ministry of Education of the Czech
Republic (project no.\ MSM 0021620835), by the grant agency of the
Charles University (grant no.\ 301311), by the Charles University in
Prague (project no.\ SVV-2013-265 301), and by the Deutsche
Akademische Austauschdienst (DAAD, project no.\ 50755689) is
gratefully acknowledged.

\appendix
\section{Details of the FEL calculation}
\label{sec:FEL_app}

Taking the sequence $5'$-TCCAG\ldots-$3'$ and its complementary part
$3'$-AGGTC\ldots-$5'$ as an example, the formation energy from
equation~(\ref{eq:gform}) reads $G_{\rm form}=\epsilon_{\rm
  TC}+\epsilon_{\rm CC}+ \epsilon_{\rm CA}+ \epsilon_{\rm AG}+\ldots$
For its calculation we use the bp energies $\epsilon_{\mu\nu}$ at $T=
25^{\rm o}$C and 1~M monovalent salt concentration, as listed in the
table~\ref{tab:NNBP} \cite{Huguet2010}.  The term $G_{\rm loop}$ in
equation~(\ref{eq:gform}) refers to the free energy reduction as a
result of the release of the end-loop. It is estimated from
\cite{SantaLucia1998,Zuker2003} as $G_{\rm loop} =1.58$~kcal/mol.

\begin{table}[b]
  \caption{Nearest-neighbour bp energies (see equation~\ref{eq:gform}) in 
    kcal/mol at $T=25^{\rm o}$C
    and 1~M monovalent salt concentration \cite{Huguet2010}.}
\begin{indented}
\lineup
\item[]
\begin{tabular}{c c c c c}
\br
 bases & $\epsilon$ & & bases & $\epsilon$ \\ \mr
 AA, TT   & -1.23 & & CC, GG & -1.93 \\
 AT  &  -1.17 & & CG & -2.37 \\
 TA & -0.84  & & GC & -2.36 \\ 
 AC, GT   & -1.49 & & AG, CT   & -1.36  \\ 
 CA, TG & -1.66 & & GA, TC & -1.47 \\ \br
\end{tabular}
\label{tab:NNBP}
\end{indented}
\end{table}

To account for the elastic response of released ssDNA, different types
of models can be assumed to calculate the mean end-to-end distance
$u_l(f)$ of the ssDNA in the force direction. The contour length of
released ssDNA from $n$~bps is
\begin{eqnarray}
l = 2dn + dn_{\rm{loop}} \delta(n,N)\,.
\label{eq:contour}
\end{eqnarray}
Here, $d=0.59$~nm/base \cite{Huguet2009} is the interphosphate
distance and $n_{\rm{loop}}$ the number of bases per loop.  The FJC
model, with an extra term suggested in \cite{Smith1996}, predicts
\begin{eqnarray}
  u_l(f)=l\left(1+\frac{f}{Y}\right)
  \left[\coth\left(\frac{bf}{k_\mathrm{B}T}\right)-
    \frac{k_\mathrm{B}T}{bf}\right]\,, 
\label{eq:FJC}
\end{eqnarray}
where $Y$ denotes the Young modulus and $b$ the Kuhn length.  Under
the working conditions for the NN model, one can use $b=1.15$~nm and
$Y= \infty$ \cite{Huguet2010}. In the WLC model
\cite{Bustamante1994a}, the elastic force for an elongation $u_l$ is
\begin{eqnarray}
  f (u_l)= \frac{k_{\rm{B}}T}{P}\left[\frac{1}{4\left(1-u_l/l \right)^2} 
    -\frac{1}{4} + \frac{u_l}{l}\right]\,,
\label{eq:WLC} 
\end{eqnarray}
and to obtain $u_l(f)$, this equation has to be inverted; the
persistence length $P$ lies in a typical range of 1.0 to 1.5~nm
\cite{Dessinges2002}.

We have chosen to model $u_l(f)$ according to the FJC model, see
equation~(\ref{eq:FJC}). Let us note that both the FJC and WLC
approaches give similar good results (for a detailed discussion, see
\cite{Huguet2010, Engel2011}).  Because $u_l(f)$ is monotonically
increasing with $f$, it has a unique inverse $f_l(u)$, which is the
force exerted by a ssDNA chain with mean end-to-end distance
$u$. Setting $\dxss=u_l(f)$, we obtain
\begin{eqnarray}
\Gstr = \int \limits_{0}^{\dxss} \textrm{d}u' \, f_l(u') = 
f\dxss - \int \limits_0^{f} \textrm{d}
f' \, u_l(f')\,.
\label{eq:Gstr}
\end{eqnarray}

\section*{References}

\providecommand{\newblock}{}

\end{document}